# Dynamic friction unraveled using an atomically defined model system

Norio Okabayashi[1]*, Thomas Frederiksen[2,3]†, Alexander Liebig[4], Franz J. Giessibl[4]‡

[1]Graduate School of Natural Science and Technology, Kanazawa University; Ishikawa, 920-1192, Japan
[2]Donostia International Physics Center (DIPC); San Sebastián, 20018, Spain
[3]IKERBASQUE, Basque Foundation for Science; Bilbao, 48013, Spain
[4]Institute of Experimental and Applied Physics, University of Regensburg; Regensburg, D-93053, Germany

*Email: okabayashi@staff.kanazawa-u.ac.jp
†Email: thomas_frederiksen@ehu.eus
‡Email: Franz.Giessibl@ur.de

Abstract: The pervasive phenomenon of friction has been studied at the nanoscale by controlled manipulation of single atoms and molecules, which permitted a precise determination of the static friction force necessary to initiate motion. However, much less is known about the atomistic dynamics during manipulation. Here we reveal the complete manipulation process of a carbon monoxide molecule on a copper surface at low temperatures using a combination of atomic force microscopy, vibrational spectroscopy for different isotope molecules and density functional theory. We measured the energy dissipation associated with manipulation and relate its origin to hysteresis involving an intermediate state, which enables an atomistic interpretation of dynamic friction. Our results show how friction forces can be controlled and optimized, facilitating new fundamental insights for tribology.

Friction is a pervasive phenomenon, yet elusive because it arises from the interplay of several physical mechanisms typically spanning very different length scales[1-4]. Seeking to isolate these, research on friction has reached the atomic level in the past decades owing to the progress of experimental techniques represented by scanning tunneling spectroscopy (STM) and atomic force microscopy (AFM)[5,6], and by computer simulations. Indeed, the empirical laws observed at the macroscopic scale have been linked to processes at the nanoscale[7-14].

The most elementary approach to study friction would be a single atom moving over a surface with STM and AFM[15-21]. These techniques are primarily used to visualize single atoms and molecules on surfaces, and even allow to manipulate them by the interaction force from the tip, enabling the fabrication of fascinating structures such as quantum corrals[22], computing devices[23] and molecular graphene[24]. Controlled manipulation of atoms with STM/AFM at liquid helium temperatures provides a powerful means to investigate friction and lubricity at the fundamental level in a highly reproducible way[19-21], because the system consists of only three bodies involving two atomic-scale contacts and thermal diffusion effects can be minimized. In addition, the geometry of the surface and tip apex[25] as well as the adsorption site of the atom can all be determined before and after the manipulation. A remarkable result obtained this way was the measurement of the lateral force needed to initiate sliding of single atoms[19], i.e. the determination of the static friction force. However, insight into dynamic friction (force needed to keep sliding) at the atomic scale requires a complete picture of the dynamics, involving intermediate states and energy dissipation, which is missing so far.

Here we address this challenge by revealing the dynamics during manipulation of a CO molecule on a copper (110) surface, probably the most widely-studied molecule in the field of surface science[26]. Compared with single atoms the benefit of this slightly more complicated system is that CO exhibits observable low-energy vibrational modes that reveal the bonding strength of the molecule to the substrate, allowing the identification of intermediate states. By utilizing AFM with a vertically oscillated tip[25] at 4.4 K, vibrational spectroscopy with STM[27,28], and density functional theory (DFT), we identify the contact point and the intermediate state in the reaction pathway responsible for energy dissipation.

When the tip is located far from the substrate, CO adsorbs on a top site in an upright configuration with its C atom bound to the Cu atom (Fig. 1a, top). By approaching the metallic tip over the top site to a close enough distance, CO moves along the [$\bar{1}$10] direction to a bridge site (Fig. 1a, bottom). Bringing a vertically oscillating tip close to the surface directly over the CO molecule leads to correlated lateral jumps of the CO molecule between top and bridge sites. When the oscillating tip approaches on a laterally shifted location closer to one of the neighboring top sites, the CO molecule is also manipulated from top to bridge site. However, in this case the CO molecule irreversibly ends up in the neighboring top site (Fig. 1b) that is closer to the metal tip apex when the tip retracts.

To first investigate this manipulation dependence on the lateral tip position ($x$), we measured the frequency shift $\Delta f$ of the vertically oscillating force sensor with a metallic tip as a function of its height $z_l$ in four characteristic situations (Figs. 1c-g, Figs. S1-2): $x/d = 0$ (top site), $\pm 0.25$ (midpoint between the top and bridge site), $\pm 0.50$ (bridge site) and $\pm 0.63$, where $d$ is the distance between nearest neighboring Cu atoms (255 pm). As shown in Fig. 1d, when the tip over the top site approaches CO (red curve), first $\Delta f$ decreases until $z_l = 160$ pm and then increases. By further approaching the tip beyond $z_l = 93$ pm (black arrow), $\Delta f$ starts to abruptly decrease, indicating that the CO has moved away from the top site. Similar abrupt decreases in $\Delta f$ are observed also for $x/d = -0.25$ and $-0.5$ (orange and green curves).

For these three cases the $\Delta f$ curves for tip retraction and tip approach are identical (Fig. S2): the CO is adsorbed on the initial top site after the retraction. However, with the tip positioned at $x/d = -0.63$ a discontinuous change in $\Delta f$ at $z_l = 125$ pm (black arrow) is observed for the tip approach (Fig. 1e). In addition, the $\Delta f$ curve for tip retraction is totally different from that for tip approach when $z_l > 125$ pm. After tip retraction, we confirmed that the CO had been manipulated to the neighboring top site that corresponds to $x/d = -1$. These observations indicate that the discontinuous change of the $\Delta f$ curve in the approach direction is caused by a lateral manipulation of the CO to the neighboring top site, while a reverse manipulation does not occur during retraction (Fig. 1b). Note that these manipulation characteristics are also observed for the lateral tip positions with positive sign ($x > 0$), (Figs. S1-S2).

To determine the adsorption geometry of CO after the abrupt decrease in $\Delta f$, inelastic electron tunneling spectroscopy (IETS)[27,28] was performed in the contact regime. This measurement (inset of Fig. 1d, black line) reveals significant differences from the conventional IETS for CO on a top site[23,28] (gray line). We further measured the vibrational energy shifts for CO molecules with different isotopes ($^{13}C^{16}O$ and $^{12}C^{18}O$) and compared the experimental results (Fig. S3) with our DFT calculations (Tables S1-S3), which allows us to conclude that CO is indeed manipulated to the bridge site.

Fig. 1f shows the experimental potential energy between tip and CO until manipulation occurs, indicating that the process takes place in the repulsive force regime. Moreover, when an abrupt decrease in $\Delta f$ is observed, a noticeable energy dissipation of a few meV per cycle of the tip oscillation arises (Fig. 1g), measured from the change in the excitation voltage to oscillate the cantilever at constant amplitude[25]. One feature of the dissipation signal is that its onset occurs at larger tip heights when the lateral tip position changes from top towards bridge (red, orange and green curves). On the contrary, in the case where the CO was manipulated to the neighboring top site, we do not see an increase of dissipation (Fig. 1g, blue line).

In order to understand the energy dissipation that occurs during these manipulation processes, we calculated the potential energy between model tip structures and a CO on Cu(110) as a function of lateral and vertical tip position ($x$ and $z_{cal}$) (Figs. S4-S6). Two exemplary cases of the lateral tip position for a relatively inert tip structure $Cu_{11}$ as defined in Fig. S5 are shown in Figs. 2a-b: $x/d = \pm 0.3, \pm 0.8$. When the tip is located laterally near the top site where the CO molecule is initially adsorbed (Fig. 2a), the crossover to molecular adsorption in the bridge configuration is energetically preferred in the repulsive force regime, which is consistent with the experimental observations (Fig. 1f). The crossover from CO top to bridge conformations occurs at a larger vertical tip position when the lateral position is shifted away from $x=0$ (Fig. S6), consistent with the experimentally observed onset of abrupt decrease in $\Delta f$ and correlated increase in dissipation (Fig. 1d and g).

To reveal the origin of the energy dissipation we analyzed the potential energy along the reaction path (Figs. S7-S8) as a function of tip-sample distance. Figure 2c shows the case when the tip is positioned close to the top site where the CO molecule is initially adsorbed ($x/d = \pm 0.3$). As discussed above, CO adsorption in the bridge configuration is energetically preferred for small tip-sample distances. However, the reaction pathway calculations (Fig. 2c) show that at the vertical tip position where the energies of the two states become comparable ($z_{cal}=125$ pm) a spontaneous transition from the CO on the top site to that on the bridge site is prevented owing to the existence of an energy barrier between CO on top and bridge conformations. Here, it is important to note that in our experiment the sensor oscillates with a peak-to-peak amplitude of 40 pm between its lower and upper turnaround points. If the tip oscillates around $z_{cal}=125$ pm, we see that the barrier disappears close to the lower turnaround point ($z_{cal}=110$ pm) and the transition of the CO molecule to the bridge site is promoted. On the contrary, when the tip retracts from the sample towards its upper turnaround point (e.g., $z_{cal}=140$ pm) CO adsorption in the top site configuration becomes energetically most favorable. Importantly, however, a small energy barrier remains between the two configurations, which indicates the lower possibility of the transition back to the top site. As in this situation, the reaction of CO for the tip approach and retract direction shows a hysteresis, which is the origin of the observed dissipation signal.

At this transition process between the two adsorption sites, the metastable conformation changes to the stable conformation, which results first in the vibrational excitation of CO and subsequent decay into the ground state by creation of electron-hole pairs in the conduction band or substrate phonons[29,30]. The estimated life time of a vibrationally excited CO on a metal surface is on the order of ps[29,30] for both of the stretching and bending modes which is negligibly short compared to the time scale of tip oscillation (19 μs). Thus it is reasonable to consider that the energy dissipation for CO to reach the ground state configuration occurs immediately after the transition.

The situation changes drastically when the tip is laterally located beyond the bridge site ($|x/d| > 0.5$). In this case, CO adsorption on the neighboring top site needs to be considered, because this conformation becomes more stable than the initial top site (Fig. 2b). With the tip far away from the surface, manipulation between the two top sites is, however, prevented owing to the energy barrier along the reaction path, whose height is approximately determined by the potential energy for the CO molecule adsorbed on the bridge site ($z_{cal}$=200 pm in Fig. 2d). When the tip approaches sufficiently close to CO, the manipulation from top to bridge is induced ($z_{cal}$=125 pm) similarly to the case for |$x/d$| < 0.5. However, when the tip retracts from the molecule, the barrier from bridge to neighboring top decreases ($z_{cal}$=160 pm), which eventually results in the manipulation to the neighboring top site. This transition occurs only one time for repeated approach and retraction of the tip, because CO on the neighboring top site is always more stable than that on the initial top site at this lateral tip position in the attractive force regime. As the CO molecule is manipulated only once, no energy dissipation can be observed in a time-averaged experiment (Fig. 1g).

The scenario described above also explains the lateral manipulation processes. Figs. 3a-d show the theoretical potential energies between the tip and CO adsorbed on three different sites (top, bridge, neighboring top) for four selected cases of tip heights (see Figs. S9a-b for full data set), where the tip initially located on the top site at $x/d$ =0 is swept towards the neighboring top site at $x/d$ =1.0. Meanwhile, CO is initially adsorbed on the top site at $x/d$=0. For simplicity, we may consider that the transition between CO in top and bridge configurations occurs spontaneously when their energies become equal. For large tip-sample distances (Fig. 3a), when the tip moves beyond the bridge site ($x/d$ >0.5), the energy of CO on the neighboring top site (blue) becomes lower than that on the top site (black). However, this manipulation is prevented owing to the presence of an energy barrier ($E_b$) as described in context of Fig. 2d. This situation is changed by lowering the tip height (Fig. 3b) where the black and red curves intersect around $x/d$~0.7: CO can now be manipulated to the bridge site, which results in a further manipulation to the neighboring top site (see Fig. 2d). When the tip-sample distance is further decreased (Fig. 3c), the black and red curves already intersect when the tip laterally approaches the bridge site at $x/d$ = 0.5. In this case, the CO molecule is first manipulated to the bridge site where it stays until the red curve intersects with the blue curve and the molecule is manipulated to the neighboring top site. Furthermore, if the tip height becomes very small (Fig. 3d), only the bridge site is available for CO, indicating that no manipulation can take place.

The aforementioned processes are summarized in Fig. 3e for the forward scan (see Fig. S10 for the backward scan). The regions depicted by gray, light blue and light red areas correspond to CO adsorbed on the top, neighboring top and bridge site, respectively. At the initial stage of manipulation, involving the transition across the blue line in Fig. 3e, we do not expect any energy dissipation as the CO molecule is manipulated only once from one top site to the next. When moving deeper into the contact regime, we predict considerable energy dissipation for manipulation across the red line, where transitions occur between top and bridge sites correlated with the vertical tip oscillation. Transitions across the red line appear initially over the bridge site and split into two lateral positions by decreasing the tip height. Indeed, as reported in Fig. 3f, these qualitative features of the dissipation onset could be experimentally resolved during CO dragging (see Figs. S11-13 for details), thus substantiating the microscopic picture of the manipulation steps. We would note that the conventional picture of manipulation based on a single hopping without intermediate sate could not reproduce this observation (Fig. S14).

Our theoretical investigations moreover provide insight into both static and dynamic friction for the manipulation. In Fig. 3b, the slope of the black line at the green cross corresponds to a static friction force $F_s$ (see Fig. S15a for details). On the other hand, the dynamic friction force $F_d$ can be evaluated by dividing the energy difference between the black and blue lines at $x/d\sim0.7$ marked as $E_d$ by the periodic distance $d$ of manipulation ($F_d = E_d/d$). We estimated the ratio of $F_d/F_s$ as a function of the tip height in Figs. S15c and found that dynamic friction ranges between 10 and 40% of static friction, consistent with the empirical law for macroscopic systems[1,2].

We have revealed the role of the intermediate state in the dynamics and energy dissipation during CO manipulation. Our approach could also be applied to similar manipulation studies such as Co on Pt(111)[19]. A next step would be to study the temperature dependence on the dissipation processes, because the thermal energy at elevated temperatures is expected to increase the transition rates and lower the energy dissipation during manipulation. This could further deepen our understanding of friction phenomena at the atomic scale.

**Acknowledgments:** The authors thank Robert W. Carpick, Enrico Gnecco, Ferdinand Huber, Takashi Kumagai, Sonia Matencio, Daniel Meuer, Hiroshi Okuyama, Magnus Paulsson, Angelo Peronio, Bo Persson, Jascha Repp, Akitoshi Shiotari, Toshiki Sugimoto, and Jay Weymouth for their valuable inputs on this research. This study was partially supported by funding from Deutsche Forschungsgemeinschaft (project no. CRC 1277, project A02), by JSPS KAKENHI Grant Number [16K04959][16KK0096][20K05320] and by Spanish AEI Grant Number PID2020-115406GB-I00. N. O. appreciates University of Regensburg for the opportunity to conduct this study as a visiting professor.

**Author contributions:** NO, TF, and FJG conceived the research. NO performed the measurements with help from AL and analyzed the data. TF conducted the DFT calculations. NO and TF wrote the manuscript. All authors discussed the results and commented on the manuscript.

**Competing interests:** FJG holds patents about the force sensor that was used in the experiments. The remaining authors declare no competing interests.

**Supplementary Materials**
Materials and Methods
Figs. S1 to S15
Tables S1 to S4
References (31–56)

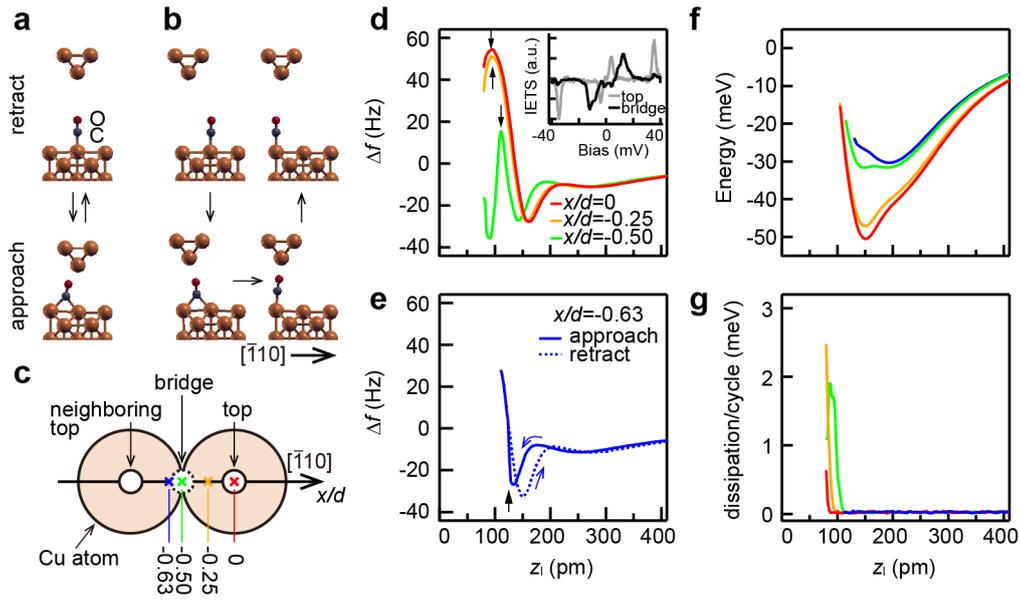

**Fig. 1. Energy dissipation processes of a CO molecule manipulated by a vertically oscillated tip.** (**a**) Schematics of the CO manipulation between top and bridge sites by a metallic tip over the top site. The thin arrows indicate the reaction processes of CO for tip approach and retract. (**b**) Same as A, but for CO manipulation from top to neighboring top site with the tip shifted laterally. (**c**) Definition of lateral tip positions along Cu [$\bar{1}$10] ($d$=255 pm): $x/d$=0 (top site), $x/d$=0.25 (midpoint between top and bridge sites), $x/d$=0.5 (bridge site), and $x/d$=-0.63. (**d**) Measured frequency shift as a function of vertical tip position ($z_l$) for three lateral tip positions, $x/d$=0, -0.25, -0.5. The oscillation amplitude $A$ was 20 pm. The inset shows typical IETS for CO in top (gray) and bridge (black) configurations with the tip located over $x/d$=0. (**e**) Frequency shift as a function of $z_l$ for the lateral tip position $x/d$=-0.63, where the force curves for tip approach and retraction are depicted by solid and dotted lines, respectively. (**f**) Measured potential energy between tip and CO up to the point of manipulation. (**g**) Energy dissipation per cycle of the vertically oscillated tip revealing the manipulation onset.

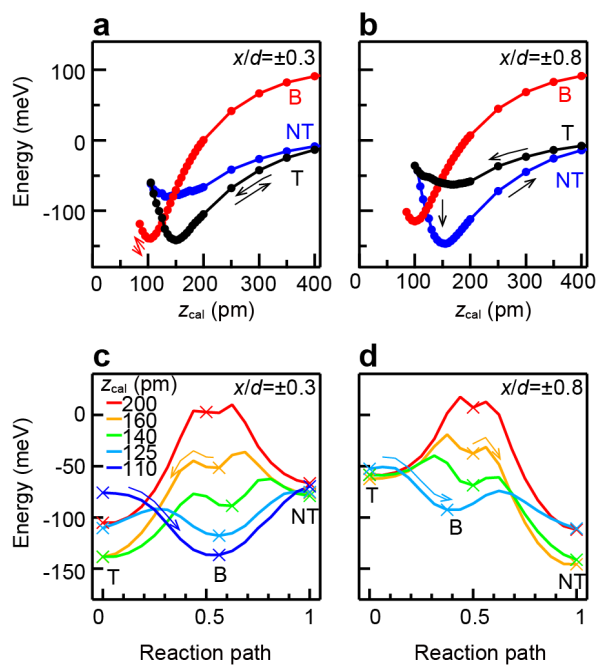

**Fig. 2. Theoretical model of CO manipulation by a vertically oscillated tip.** (**a**) Potential energy between a $Cu_{11}$ tip as defined in Fig. S4 on $x/d=\pm0.3$ and CO as a function of the vertical tip position ($z_{cal}$). The cases for CO on top (T), bridge (B) and neighboring top (NT) sites are depicted by black, red and blue lines, respectively. (**b**) Same as A but for the tip over $x/d=\pm0.8$. (**c**) Calculated reaction pathways for CO from top site (0.0) to next top site (1.0) via bridge configuration for fixed $Cu_{11}$ tip over $x/d=\pm0.3$. The cross marks correspond to CO in top, bridge, and neighboring top configurations, respectively. (**d**) Same as C but for the tip over $x/d=\pm0.8$.

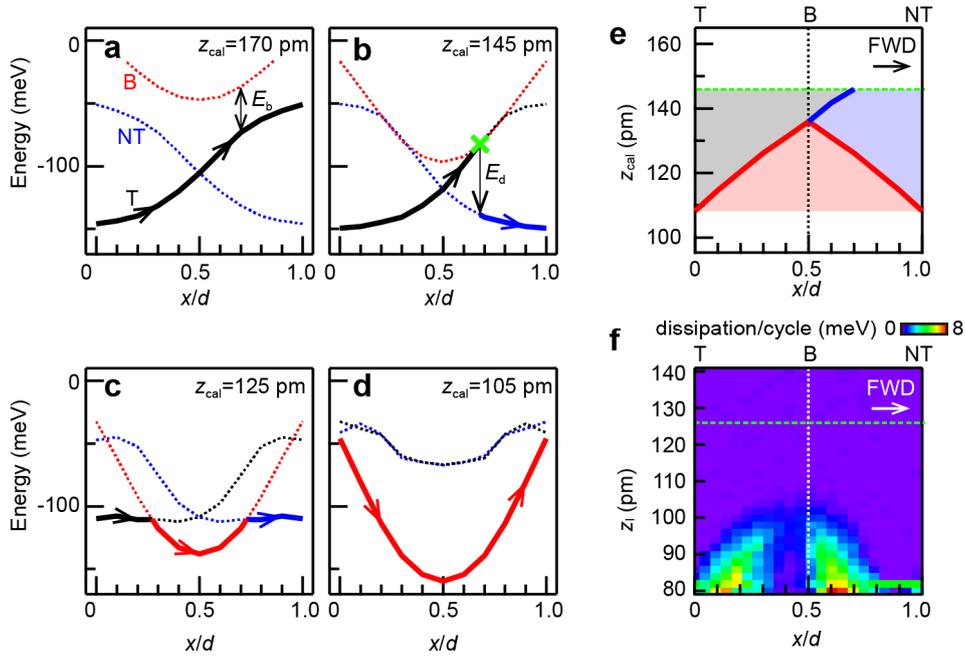

**Fig. 3. CO dragging process with energy dissipation.** (**a**) Calculated potential energy between a $Cu_{11}$ tip as defined in Fig. S4 and CO as a function of lateral tip position (at constant tip height of 170 pm) for CO in top, bridge and neighboring top configurations. The thick path represents CO configuration during the lateral tip scan. (**b-d**) Same as A but for tip heights 145 pm (b), 125 pm (c), and 105 pm (d). (**e**) Calculated CO configuration as a function of tip height and lateral position during forward tip scan. The gray, light red and light blue regions represent CO in top, bridge, and neighboring top sites, respectively. The thick lines indicate the transition points. (**f**) Experimentally detected energy dissipation per cycle of the vertically oscillated tip, acquired during lateral CO dragging. In (e) and (f), the onset of dragging is indicated by the dashed green line.

# Supplementary Materials for

## Dynamic friction unraveled using an atomically defined model system

Norio Okabayashi[1]*, Thomas Frederiksen[2,3]†, Alexander Liebig[4], Franz J. Giessibl[4]‡

Correspondence to: *Email: okabayashi@staff.kanazawa-u.ac.jp,
†Email: thomas_frederiksen@ehu.eus, ‡Email: Franz.Giessibl@ur.de

**This PDF file includes:**

Materials and Methods

Figs. S1 to S15

Tables S1 to S4

References (31–53)

**Materials and Methods**

**Experimental methods.** All measurements were performed at a low temperature (4.4 K) and under ultrahigh vacuum condition using a combined STM and AFM system (LT-SPM by ScientaOmicron GmbH) at the University of Regensburg. The sample substrate was a Cu(110) crystal, which was cleaned by repeated sputtering and annealing. CO molecules were adsorbed on the surface at low temperature with a coverage of a few molecules per 10 nm × 10 nm area.

The force field was measured using a qPlus sensor[25], where a metallic tip made from tungsten wire with a diameter of 50 μm was attached to the end of the cantilever of the sensor. The parameters of the sensor are: eigenfrequency $f_0$=52,194 Hz, stiffness $k$=1,800 N/m and quality factor $Q$ = 595,000. To measure the force and the potential energy between the tip and molecule, the frequency shift of the vertically oscillated force sensor was measured at a constant amplitude of $A$=20 pm, which was then converted to a force value and a potential energy using deconvolution methods[31-33]. In FM-AFM, the frequency shift of the sensor from its unperturbed resonance frequency is a measure of the vertical tip force gradient $k_{ts}$ averaged over the sensor oscillation as $\Delta f = f_0 <k_{ts}>/2k$ [32]. During the measurement of the frequency shift, the excitation voltage ($V_{exc}$) added to the AFM oscillation electrode to keep a constant amplitude was simultaneously measured, which was used to estimate the dissipation energy ($E_{dis}$) per cycle of the oscillation by the following equation $E_{dis}=2\pi kA^2/(2Q)\times V_{exc}/V_{exc0}$ [25], where $V_{exc0}$ is the excitation voltage when the tip is far away. The tip height $z$=0 is chosen at the point contact, where the tunneling conductance would reach the conductance quantum. When the tip is oscillated, the distance between the lower turnaround point and $z$=0 is defined as the tip height $z_l$.

The tip attached to the cantilever was also used to measure the electron tunneling current, by which the IETS[27,28] curve was measured. A modulation voltage of $V_{mod}$=1.0 mV$_{rms}$ was added to the sample bias and the second harmonic signal of the tunneling current was measured using a lock-in amplifier. We adopted radio frequency (RF) filters to attenuate the RF noise from the environment, thus increasing the resolution of the IETS measurement[34].

For all AFM and STM-IETS measurements, tips whose apices consisted of a single atom [25,35] were used (Fig. S1), because (I) these tips can exert a stronger attractive force[36], which is preferable to induce the lateral manipulation[20] and (II) these tips can provide stronger IETS signals[37].

**DFT calculations.** We computed the potential energy landscape, forces, and vibrational frequencies using periodic, plane-wave density functional theory (DFT) calculations as implemented in VASP[38,39]. To correctly account for the preferred top site adsorption of CO on Cu(110), see Table S4, we employed the vdW-DF2 nonlocal exchange-correlation functional[40-43]. Calculations were performed with the planewave energy cutoff set to 600 eV, a 4 × 4

Monkhorst-Pack *k*-point mesh, and first-order Methfessel-Paxton occupations with 0.1 eV smearing. The lattice constant for Cu was set to $a_0$=375 pm as computed with the vdW-DF2 functional. As shown in Table S4 and its caption, the choice of the exchange-correlation functional shows the trade-off relationship regarding the lattice constant and stable adsorption site.

The CO-Cu(110) system was represented in a 2 × 3 surface unit cell slab with 8 atomic layers and ~2 nm vacuum region between periodic images. Two models were explored for the tip apex geometry: a relatively inert $Cu_{11}$ cluster and a more reactive $Cu_5$ cluster (Fig. S4) with the relative coordinates fixed to those of the isolated cluster. The nominal height of the tip apex atom ($z_{cal}$) is measured from the point which is higher than the first layer of the copper substrate by 380 pm, which is close to the lattice constant: the definition is therefore similar to the experimental one. The geometry and total energy of the system were determined by relaxing CO and the topmost Cu layers for different top positions (fixed super-cell size) until residual forces on CO and the Cu surface layer were within $10^{-6}$ eV/pm. The Cu surface atoms away from the molecule were laterally constrained to avoid potential sliding effects of the top layer due to periodic boundary conditions. Background subtraction of the interaction energy was performed by evaluating also the total energy for supercells without CO. Harmonic vibrational energies were computed by finite displacements with an amplitude of 5 pm, which provides good numerical accuracy to describe the low-frequency modes. The nudged elastic band (NEB) method[44] was used to compute reaction pathways, employing a two-stage approach with three intermediate images between end points at each stage.

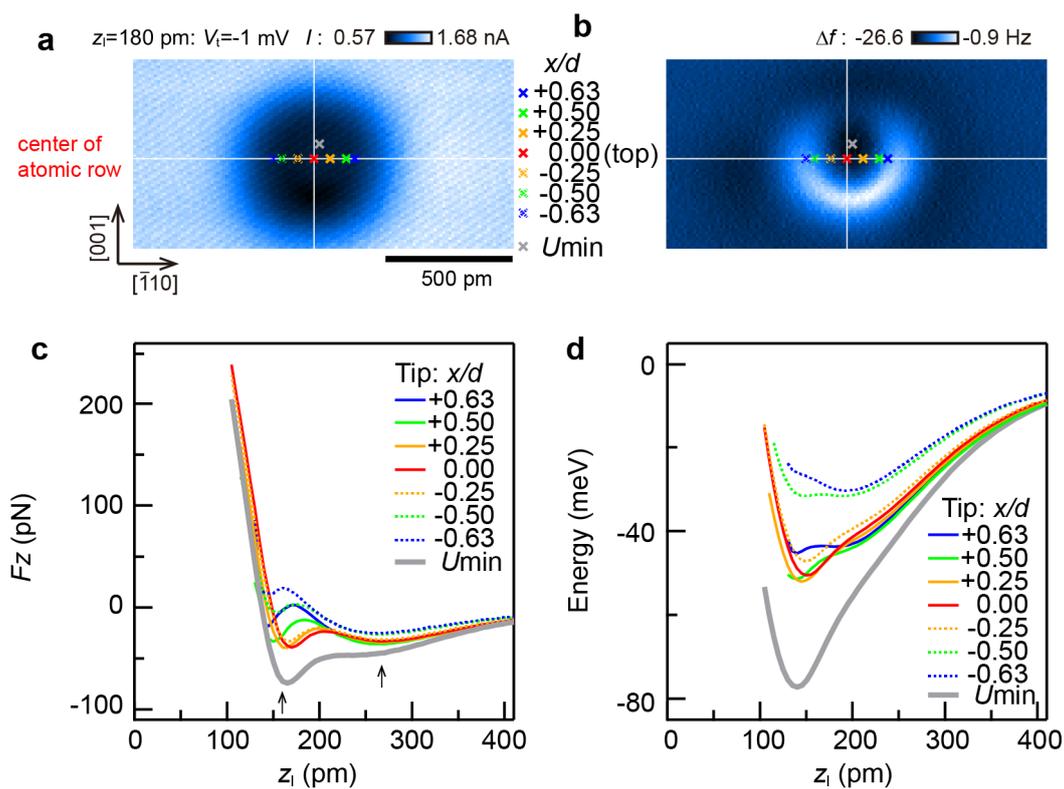

**Fig. S1. Characterization of the tip apex and tip position.** (**a**) Constant height current image of a CO molecule on a Cu(110) surface, that was measured at a sample bias of -1 mV and at a tip height of $z_l$=180 pm. The top site is determined to be the center of the depression in the image. (**b**) Constant height $\Delta f$ image above the CO acquired simultaneously with (a), which indicates that the tip apex consists of a single atom[25,35]. The lateral tip position that shows a minimum of $\Delta f$, referred to as $U_{min}$ (gray cross mark), is slightly shifted from the top site (red cross mark). (**c**) Vertical force and (**d**) potential energy for the tip on the potential minimum are shown by the thick gray lines including the cases of various lateral tip positions along the center of the atomic row (Fig. S2). In the force curve, we see two components of the attractive force marked by the arrows, which corresponds to the van der Waals interaction and the chemical interaction[45].

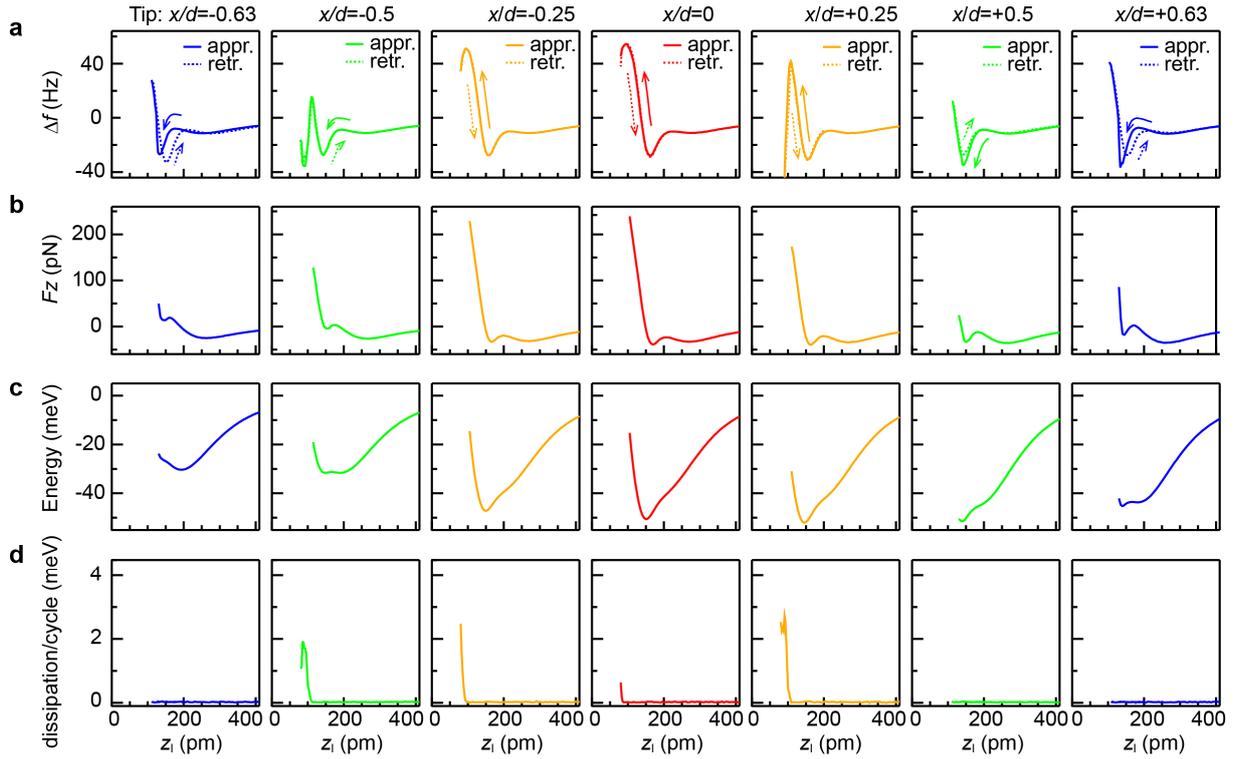

**Fig. S2. Full data set of the lateral tip position dependent manipulation processes.** (**a**) Frequency shift, (**b**) vertical force, (**c**) potential energy and (**d**) dissipation/cycle. The following cases of the lateral tip positions were investigated: $x/d =0, \pm 0.25, \pm 0.5$ and $\pm 0.63$, where the data for the negative tip position ($x/d < 0$) is identical to that shown in Fig. 1. In the $\Delta f$ curves for each tip position, the cases for tip approaching and retracting are depicted by a solid and dotted line, respectively. For $x/d = \pm 0.63$ and for $x/d =+0.5$, we see a hysteresis in the $\Delta f$ curve between the tip approaching and retracting as the CO is manipulated from one top site to the neighboring top site. On the other hand, for the lateral tip positions of $x/d =0, \pm 0.25, -0.5$, the $\Delta f$ curves between the tip approaching and retracting are identical even after the manipulation of the CO molecule between the top site and the bridge site. In the figures for the vertical forces and the potential energies, the curves are depicted for the tip approaching until the manipulations occur. In the figures for the dissipation, the curve in each panel is depicted for the tip approaching until the lowest vertical tip position in the panel (a).

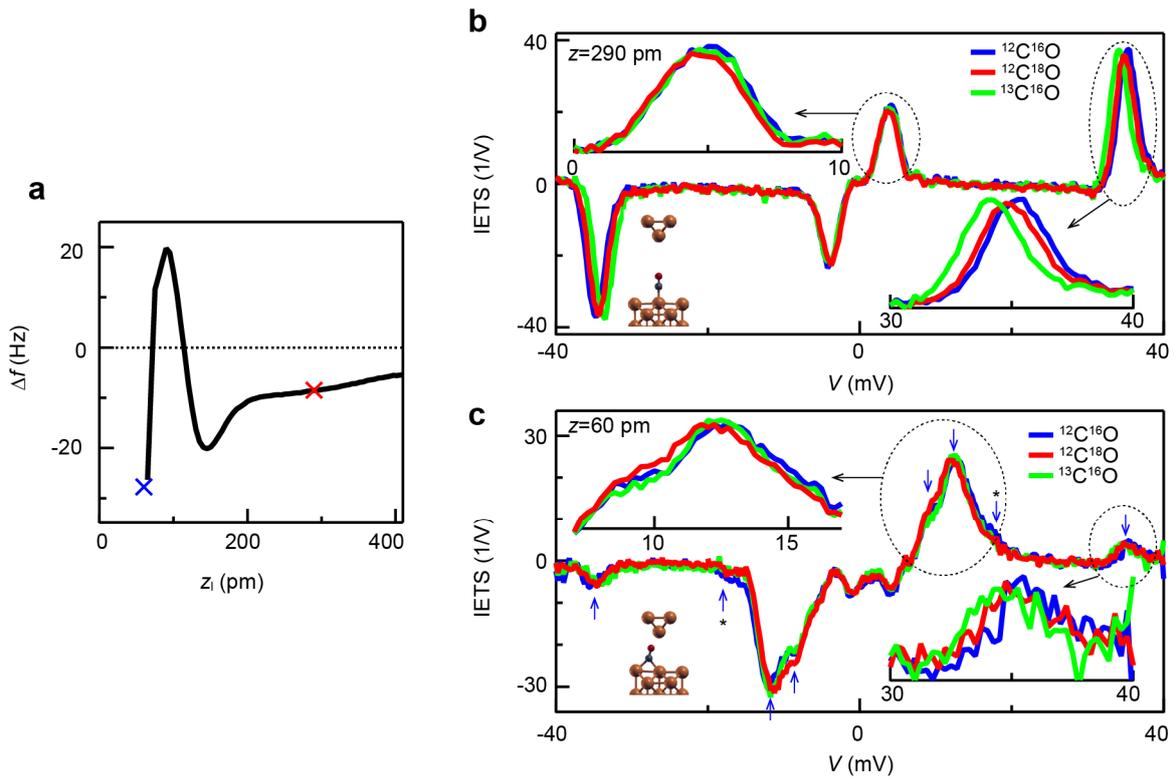

**Fig. S3. Altered adsorption site of CO from top site to bridge site.** (**a**) Frequency shift curve for a tip used for this measurement. IETS was measured at the two vertical tip positions shown by the cross marks. (**b**) Isotope dependent IETS for CO on a top site. The vertical tip position was $z=290$ pm (red cross mark in a). The two insets show the expanded IETS around the frustrated translational (FT) mode positive peak (left) and frustrated rotational (FR) mode positive peak (right). In the FR mode of CO on a top site, the C atom is more displaced than the O atom[28], thus a strong isotope shift is observed for $^{13}C^{16}O$. The case of the FT mode of CO is opposite: The O atom is more displaced, which results in a stronger isotope shift for $^{12}C^{18}O$. See Table S1 for more details. (**c**) Isotope dependent IETS for CO on a bridge site. The vertical tip position was $z=60$ pm (blue cross mark in a). We see three clear peaks at 8, 12, and 35 mV and one obscure peak at 18 mV. The two insets show the expanded IETS for the region of +7-17 mV (left) and +30-40 mV (right). Comparing to our DFT calculations for CO on the bridge site (Tables. S2-3), we tentatively assign the IETS peaks observed at 8, 12, 18 and 35 mV to FT mode along [001] and [$\bar{1}$10], FR mode along [$\bar{1}$10] and [001], respectively. According to our DFT calculation, the displacements of C and O in the FT and FR modes on the bridge site is similar to the case of CO on the top site: O is more displaced in the FT modes while C is more displaced in the FR modes. The observed significant isotope shifts by O substitution for the peak at 12 mV and by C substitution for the peak at 35 mV are consistent with these theoretical expectations. The drastic change in IETS for a very small $z$ was also reported for CO/Cu(111)[46].

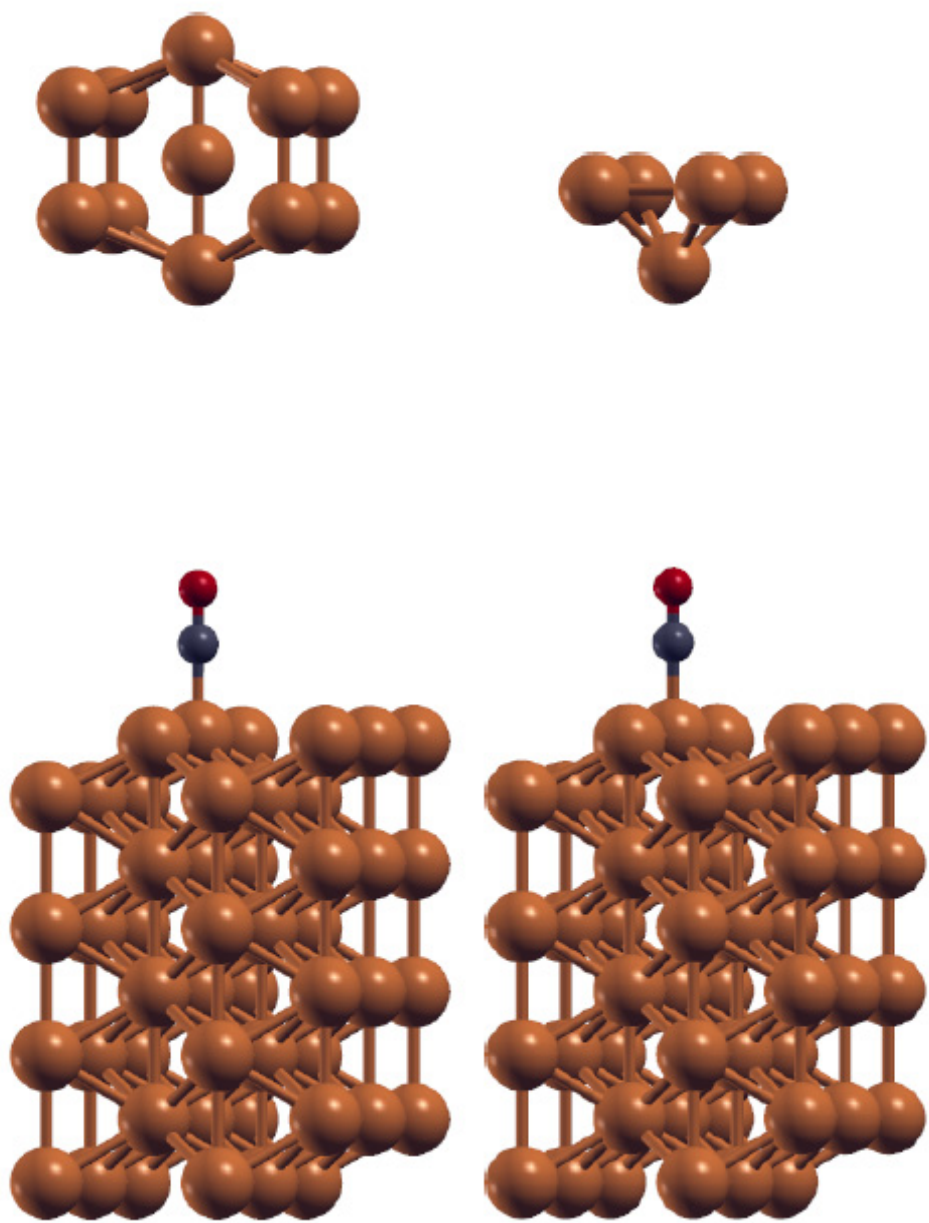

**Fig. S4. Two tip structures over CO/Cu(110).** Left: $Cu_{11}$ tip. Right: $Cu_5$ tip. As shown in Fig. S6, the Cu5 tip is about two times as reactive as the Cu11 tip.

**a** Cu$_{11}$ Tip

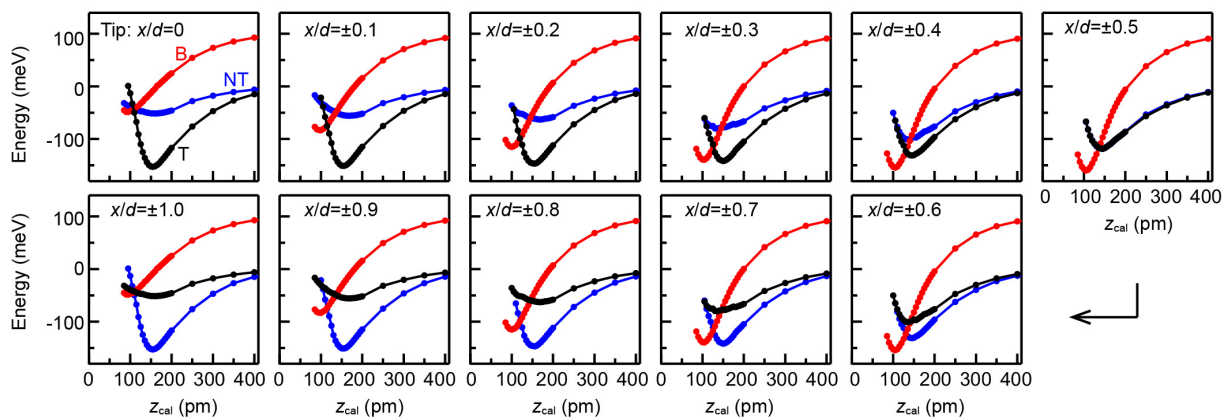

**b** Cu$_5$ Tip

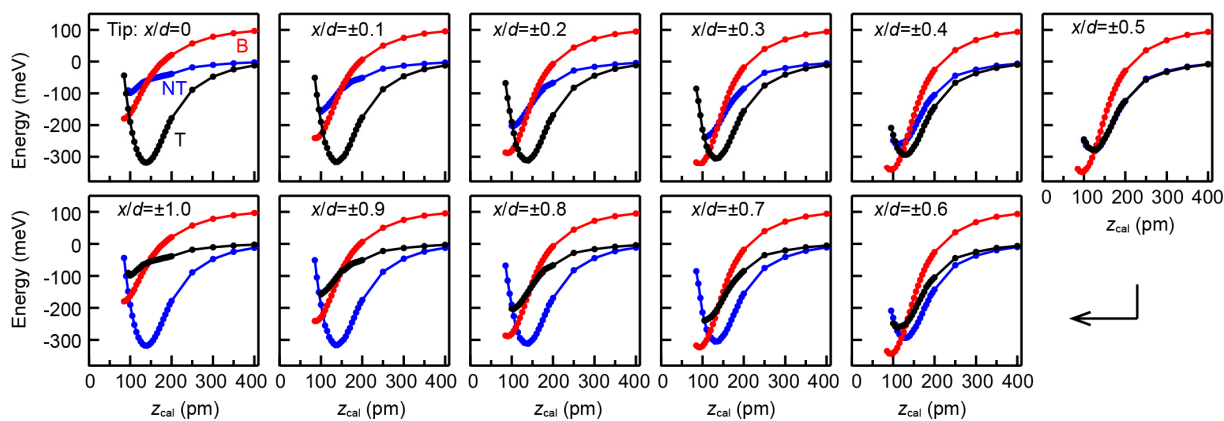

**Fig. S5. Theoretical investigation on the tip height dependent potential energies between a CO molecule on a Cu(110) surface and a Cu tip.** (**a**) The cases for the inert Cu$_{11}$ tip are shown, where the lateral tip position is systematically changed. In each panel, the black, red and blue lines represent the adsorption of CO on the top (T), bridge (B) and neighboring top (NT) site, respectively. The potential energies between t and nt are symmetric with respect to the tip position at the bridge site, e.g., the potential energy for t at $x/d$=0.2 is identical to that for nt at $x/d$=0.8. (**b**) The same as (a) but for the case of the reactive Cu$_5$ tip.

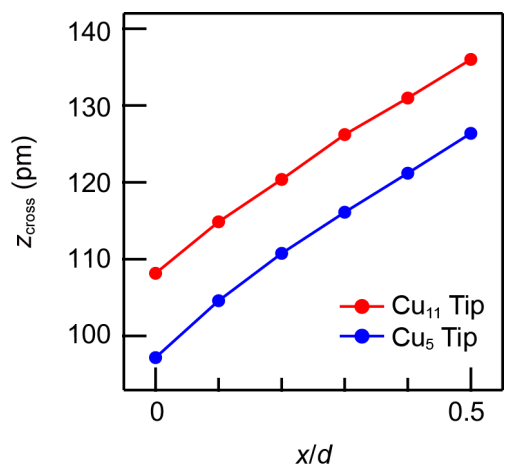

**Fig. S6. Vertical tip height where the potential energies for CO on a top and on a bridge site in Fig. S5 become identical are plotted as a function of lateral tip position.** We see that the height of the cross-point increases with moving the lateral tip position from top ($x/d$=0) to bridge site ($x/d$=0.5), which is consistent with the experimental observations in Fig. 1d and g.

**Fig. S7. Calculated nudged elastic band (NEB, see supplementary text) reaction pathways for CO from top site (0.0) to next top site (1.0) via bridge site for fixed Cu$_{11}$ tip.** The lateral tip position is systematically changed from $x/d$=0 to 1.0. In each panel, the most left and right points highlighted by the cross marks correspond to the CO on the top (T) site and that on the neighboring top (NT) site, respectively. The cross marks between these two points correspond to the CO on the bridge (B) site, which is a good indicator for the transition state when the tip is far from the molecule.

**Fig. 8. Calculated nudged elastic band (NEB, see supplementary text) reaction pathways for CO from top site (0.0) to next top site (1.0) via bridge site for fixed Cu$_5$ tip.** The same as Fig. S7, but for the Cu$_5$ tip.

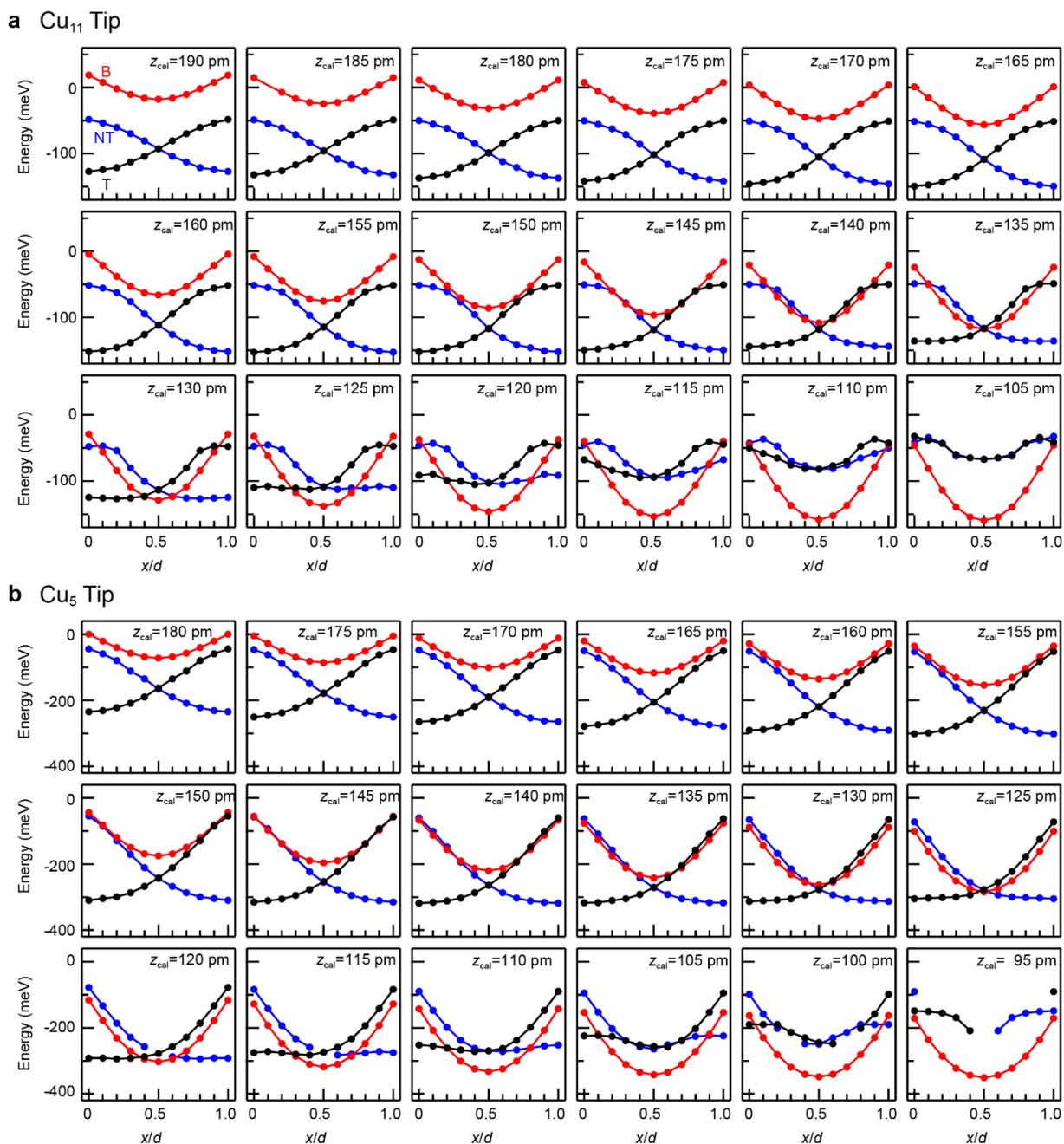

**Fig. S9. Calculated energy profiles for lateral manipulation.** (**a**) Potential energies between the inert $Cu_{11}$ tip and CO on the three adsorption sites are plotted as a function of the lateral tip position, where the tip height is systematically changed. In each panel, the black, red and blue lines represent CO on top (T), bridge (B) and neighboring top (NT) site, respectively. (**b**) The same as (a), but for the reactive $Cu_5$ tip. The potential energies for the three adsorption sites become similar for $z_{cal}$=135 pm for the $Cu_{11}$ tip and $z_{cal}$=125-130 pm for the $Cu_5$ tip, where we expect the smallest friction to laterally manipulate CO.

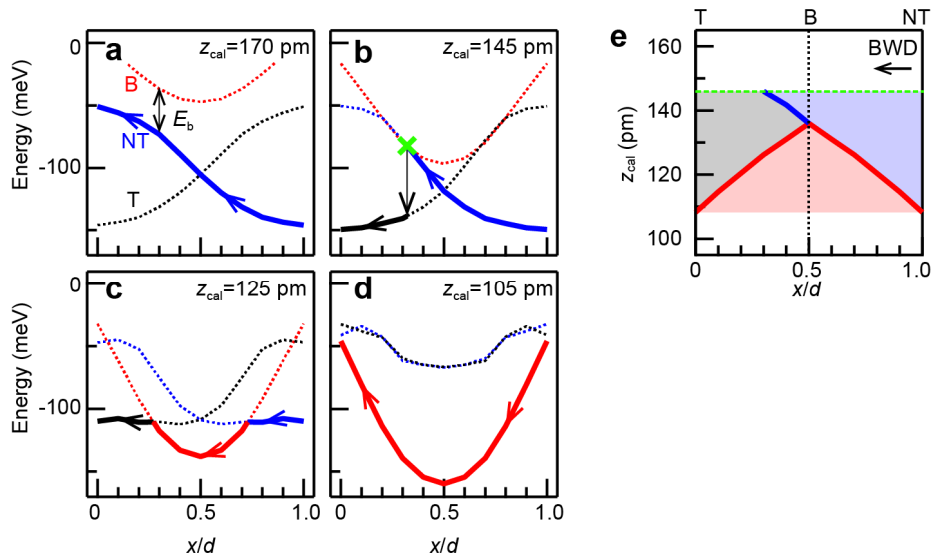

**Fig. S10. Theoretical investigation on the CO dragging process for the backward scan.** (a-d) The potential energies between the tip and CO adsorbed on top (black), bridge (red) and neighboring top (blue) sites are plotted as a function of the lateral tip position. The tip initially on the neighboring top site at $x/d$=1.0 is swept towards the top site at $x/d$=0, whereas the CO molecule is initially adsorbed on the neighboring top site ($x/d$=1.0). We may consider the spontaneous transition of the CO molecule when the energy for the neighboring top site becomes identical to that for the bridges site, similarly to the case for the forward scan described in the main text. The thick path represents CO configuration for the backward tip scan. (a) The case for the large tip-sample distance. When the tip moves beyond the bridge site ($x/d$=0.5), the energy of CO on the top site (black) becomes lower than that on the neighboring top site (blue). However, the manipulation is prevented owing to the presence of an energy barrier ($E_b$). (b) Decreasing the tip height changes the situation. The energy for the neighboring top site becomes identical to that for the bridge at $x/d$ ~0.3, where the manipulation to the bridge site occurs first, which results in a further manipulation to the top site. (c) The case for further deceasing the tip height. When the tip approaches to the bridge site at $x/d$=0.5, the energy for the bridge site (red) is already lower than that for the neighboring top site (blue). In this case, the manipulation to the bridge site occurs first, where the CO molecule keeps adsorbing on the bridge site until the energy for the bridge site (red) becomes identical to the that for the top site (black). (d) The case for the very low tip height, where the CO molecule on the bridge site is always stable: no manipulation can take place. (**e**) The dragging process for the backward scan is summarized, where the onset of the dragging is indicated by the green dotted line. The regions depicted by gray, light blue and light red correspond to the CO molecule on the top, neighboring top and bridge site, respectively. The feature of this dragging process is basically identical to that for the forward scan (Fig. 3e), however, the slope of the blue line is opposite between the forward and backward scans: at the beginning of the dragging, the lateral tip position where the manipulation occurs is slightly different between the forward and backward scan, which becomes smaller by further decreasing the tip height. This trend is consistent with the experimentally observed asymmetries with scan direction (see Fig. S12, $z_l$=124-122 pm). When the tip is swept for a larger scan range, the repetition of Figs. 3e and S10e is expected.

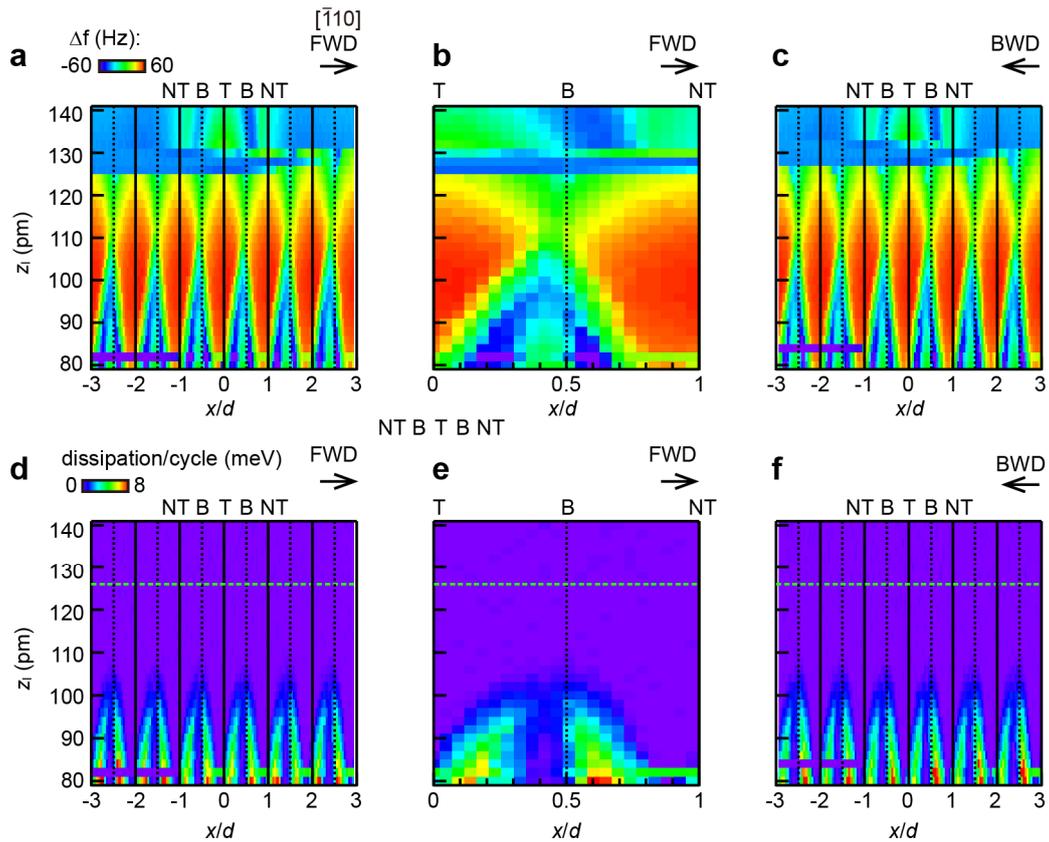

**Fig. S11. Energy dissipation during dragging of a CO molecule.** (**a**) Frequency shift data as a function of the lateral ($x$) and vertical ($z_l$) tip positions, where the top/neighboring top (T/NT) and bridge (B) sites are depicted by the solid and dotted lines, respectively. (**b**) An enlarged view of (A) excerpted from $x/d$=0 to 1. (**c**) The same as (a), but for the backward scan. In (a) and (c), CO was initially adsorbed on the top site at $x/d$=0, and the tip was set at $z_l$=140 pm and $x/d$=-3. The tip was first swept forward, then backward. After finishing one lateral scan, the tip height was decreased by 2 pm and the lateral scan was repeated. At the initial stage of this measurement, CO stays adsorbed in the initial top site. Then at the backward scan at $z_l$=132 pm, the manipulation to the neighboring top site on the right occurs (see also Fig. S12a). By further decreasing the tip height, for a while, the CO remains at $x/d$=3, because the slightly asymmetric tip apex in this case favors manipulation to the right. Then, when the tip height reaches to $z_l$=126 pm manipulation to the left side starts to occur, which results in the so-called dragging of the molecule where the molecule is manipulated along the $[\bar{1}10]$ direction as if trapped by the tip. For a very small tip height ($z_l$=82 pm) dragging eventually fails. (**d**) Energy dissipation per cycle of the vertically oscillated tip, which was acquired simultaneously with (a). (**e**) An enlarged view of (d) excerpted from $x/d$=0 to 1. The image is identical to Fig. 3f. (**f**) The same as (d), but for the case of backward scan. At the initial stage of the dragging from $z_l$ = 126 (green dotted line) to 110 pm, CO is manipulated without a signal in the energy dissipation (see also Fig. S12b). However, when the tip height reaches $z_l$ = 108 pm, the onset

of energy dissipation is observed for the tip over the bridge sites. By further decreasing the tip, the peak located on the bridge sites is split into two lateral tip positions where the dissipation energy increases. These features of the CO dragging with dissipation were also confirmed in constant height raster scan images (Fig. S13).

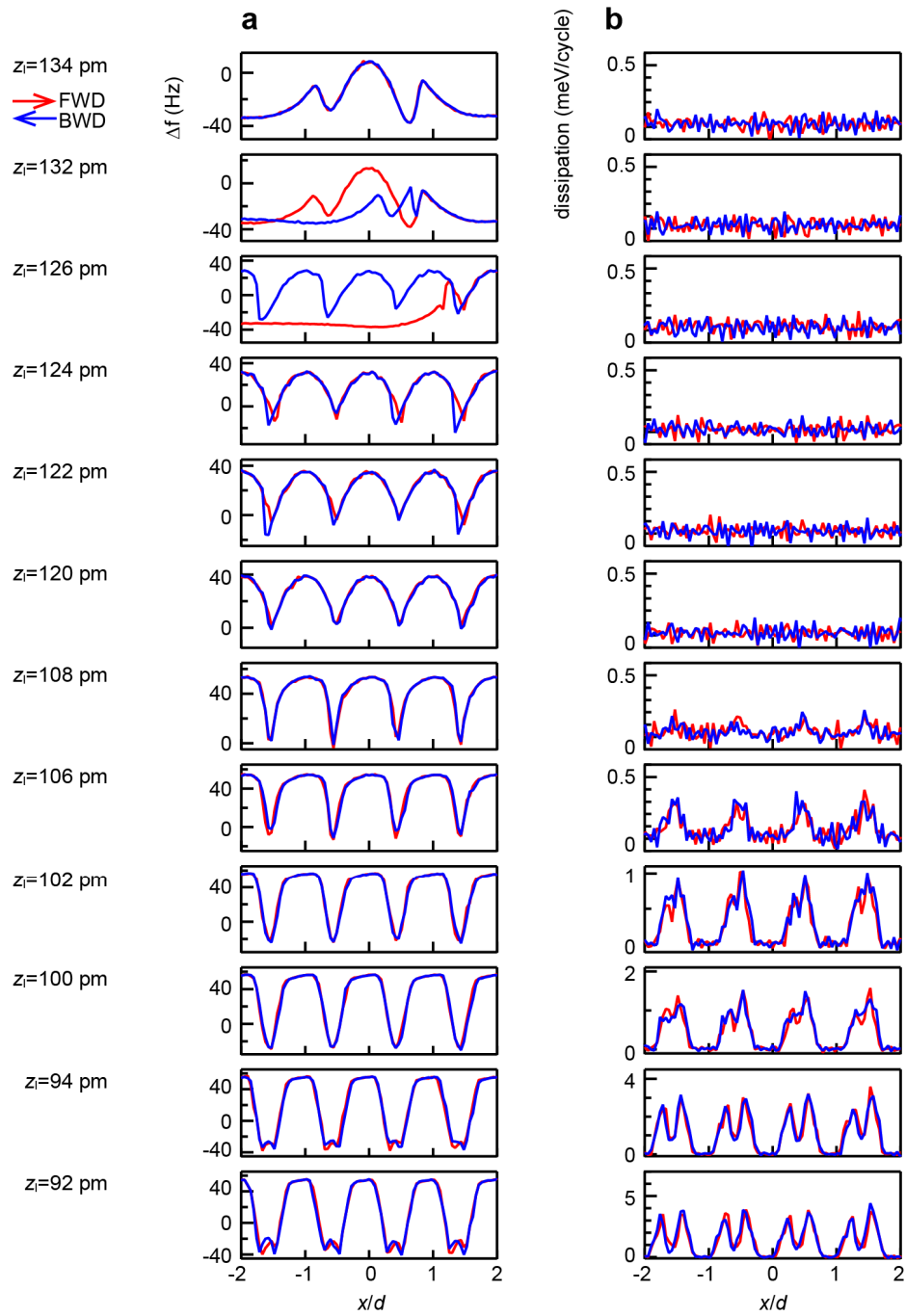

**Fig. S12. The detail of the frequency shift and energy dissipation during CO dragging.** (**a**) Selected line scans of the frequency shift measurement in Fig. S11 are shown for both the forward and backward scan directions. In Fig. S11, the scan range is from $x/d$=-3 to 3, however, the range from $x/d$=-2 to 2 was excerpted here. (**b**) The same as (a), but for the energy dissipation.

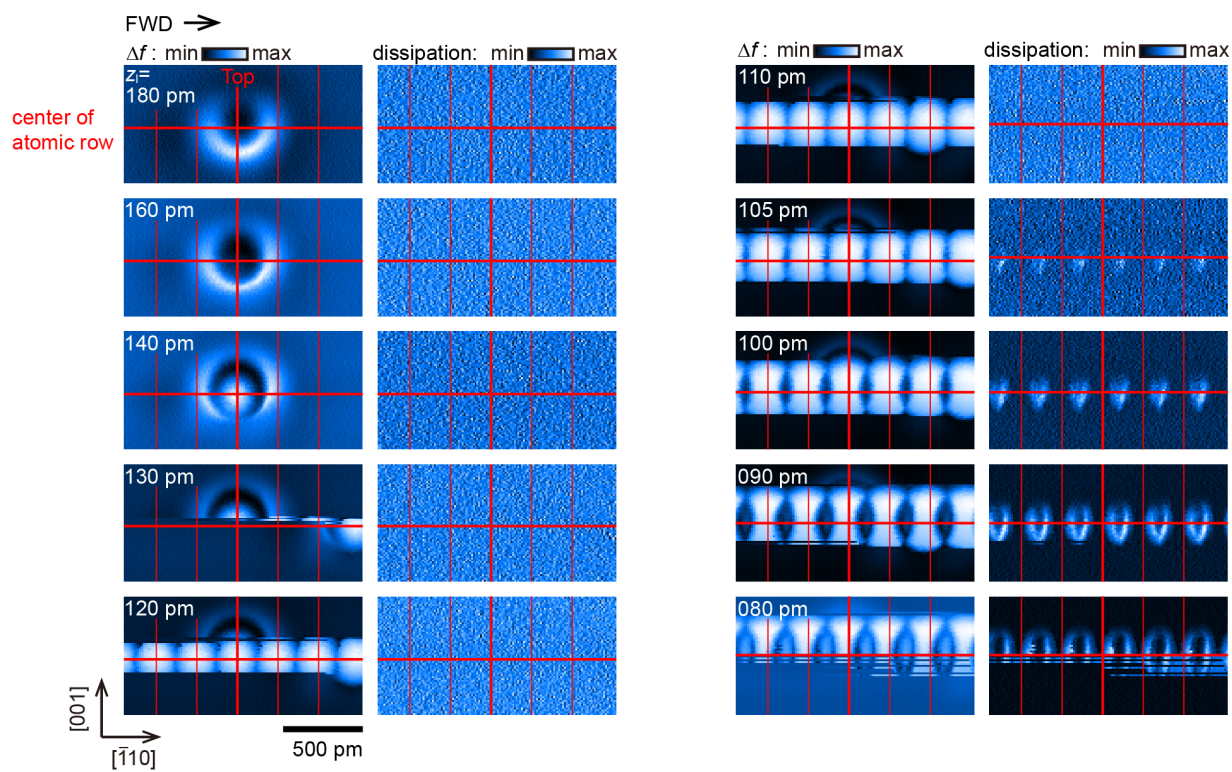

**Fig. S13. Frequency shift and energy dissipation images in constant height raster scans.** In each image, the tip is initially located on the upper left corner of the image and the CO molecule is initially adsorbed on the top site indicated by the cross-point of the red thick lines. The fast-scan direction is horizontal and the slow-scan direction is from up to bottom. Here, only the forward scan direction is shown. The cross-points of the vertical and lateral red lines mark the positions of the atoms along one atomic row. When the vertical tip position is relatively large ($z_l$ =180 pm), an attractive feature (decrease in $\Delta f$) appears over CO. By further approaching the tip, this attractive feature changes to a repulsive feature, and finally at $z_l$ =130 pm manipulation to the neighboring top site on the right side occurs. Note that in the case of the present tip, the slight asymmetry of the tip promotes manipulation to the right side, which eventually results in the CO being located on the rightmost position. When $z_l$ decreased to 120 pm, the manipulation to the left side also occurs, resulting in the so-called dragging, where CO can be manipulated along the [$\bar{1}$10] direction as if trapped by the tip. At the initial stage of the dragging, we do not see a dissipation signal. However, at $z_l$ =105 pm, the onset of the dissipation appears at the bridge sites. By lowering the tip height, the dissipation signal at the bridge sites is split into the two peaks which are symmetric with respect to the bridge site.

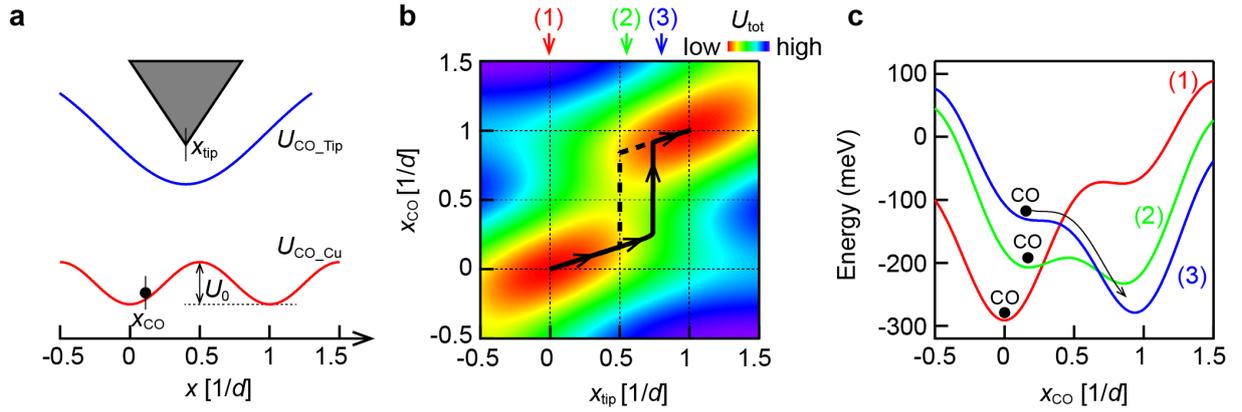

**Fig. S14. Prandtl and Tomlinson model[2] to qualitatively explain the energy dissipation process during a manipulation, where only a single hopping process is considered.** (**a**) We consider the situation where a CO molecule moves on a one dimensional sinusoidal potential along the $x$ axis (red line): $U_{CO\_Cu}=-U_0/2\times\cos(2\pi x_{co}/d)+U_0/2$, where $U_0$ (=97 meV [47]) is the activation energy for the manipulation, $x_{co}$ is the CO position and $d$ is the distance between the neighboring top sites. We also consider that the tip is scanned along the $x$ axis and the interaction between CO and the tip can be expressed by a Gaussian (blue line): $U_{CO\_tip}=-E_0\times\exp[-\{(x_{CO}-x_{tip})/wd\}^2]$, where $E_0$ is the maximum interaction potential, $x_{tip}$ is the tip position, $w$ is the width of the potential here temporarily adopted to be 0.8 from our DFT calculation. (**b**) The total potential energy $U_{tot}=U_{CO\_Cu}+U_{CO\_tip}$ for the case of $E_0=3U_0$ is plotted as a function of $x_{co}$ and $x_{tip}$. The dashed line corresponds to the minimum energy for the tip moving from the initial top site ($x_{tip}/d=0$) to the neighboring top site ($x_{tip}/d=1$). The solid line is the actual trajectory of the CO molecule for the tip scan at 0 K. (**c**) $U_{tot}$ for the three tip positions: $x_{tip}/d=0$ (red), 0.55 (green), 0.8 (blue). When the tip is located on the initial top site ($x_{tip}/d=0$), this top site is most stable for CO. When the tip is located beyond the bridge site ($x_{tip}/d=0.55$), the most stable site is changed to the neighboring top site. However, manipulation does not occur at 0 K, as an energy barrier of about 15 meV in green line (2) would still need to be overcome. When the tip moves further towards the neighboring top site ($x_{tip}/d=0.8$) shown by the blue line (3), the barrier disappears resulting in the CO manipulation, where we can expect an energy dissipation. This model nicely illustrates the crucial process of friction: stick-slip motion. However, in order to correctly understand the process occurring at the actual system, ab-initio calculations are mandatory, because the actual potentials between CO, Cu surface and tip are more complicated than the simple sinusoidal for $U_{CO\_Cu}$ and the Gaussian for $U_{CO\_tip}$ (Figs. S5, S7-9). For example, if we would not consider the additional adsorption site at the bridge site explained in the main manuscript, we could not interpret the precise shape of the energy dissipation observed in the experiment. Note that the similar tip induced change of the potential energy landscape of a single molecule adsorbed on a surface was also reported for different systems[48-50].

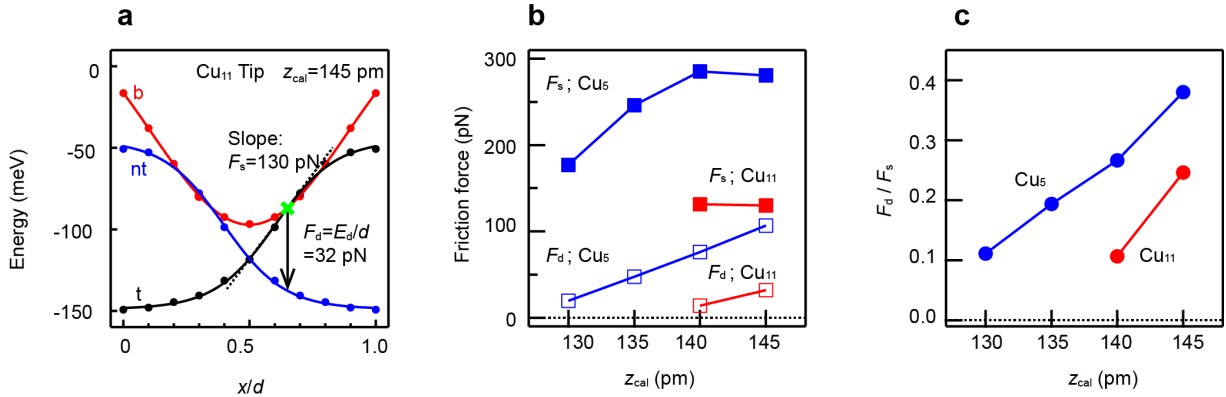

**Fig. S15. Static friction and dynamic friction.** (**a**) One of the frames of the calculated energy profiles for lateral manipulation used in Fig. S9: $Cu_{11}$ tip and $z_{cal}$=145 pm. The black and blue lines are obtained by fitting the calculated data points with sigmoid functions, $y=b + m/[1+\exp\{(x_0-x)/rd\}]$, where $b$, $m$, $x_0$ and $r$ are the fitting coefficients, $y$ is the dependent variable and $x$ is the independent variable. The red line is obtained by a fit with a Gaussian function. Manipulation from the top site to the neighboring top site is considered to occur when the black line intersects with the red line as shown by the green cross mark. The slope of the black line at this intersection (dotted line) corresponds to the lateral force needed to manipulate CO[19,20], i.e., the static friction ($F_s$). On the other hand, the energy difference between the black and blue line at this intersection corresponds to the energy dissipation ($E_d$) as shown by the black arrow, which is equivalent to the work needed to manipulate CO. The dynamic friction ($F_d$) is acquired by dividing the energy dissipation by the periodic distance ($d$) for the manipulation (255 pm). (**b**) Static and dynamic friction as a function of the vertical tip-sample distance estimated from the data shown in Fig. S9. (**c**) The ratio of dynamic friction to static friction is plotted as a function of the vertical tip position for the two tip models. The ratio ranges between 10-40%, which is consistent with the empirical law for macroscopic systems[1,2].

**Table S1. Experimental investigation on the low-energy vibrational energy shifts for an isotope substituted CO molecule**. (**a**) The case for CO on top site, where the perturbation from the tip to the molecule is negligibly small ($z=290$ pm)[28]. The numbers in the table represent the vibrational energies of CO (in parenthesis, the isotope shift relative to the normal molecule). (**b**) The same as (a), but for CO adsorbed on bridge site, where the vibrational energies are strongly influenced by the perturbation from the tip. The peak heights at 8 and 18 meV are so small that it is difficult to discuss the isotope dependence. The relative vibrational energy shifts in the experiments are consistent with our theoretical calculations (Tables S2-3). FR: Frustrated rotation, FT: Frustrated translation.

**a**, CO on top site: $z=290$ pm

|    | $^{12}C^{16}O$ (meV) | $^{13}C^{16}O$ (meV) | $^{12}C^{18}O$ (meV) |
|----|---|---|---|
| FR | 34.97 | 33.95 (-3.0%) | 34.63 (-1.0%) |
| FT | 4.02  | 3.99 (-0.7%)  | 3.79 (-5.7%)  |

**b**, CO on bridge site: $z=60$ pm

|    | $^{12}C^{16}O$ (meV) | $^{13}C^{16}O$ (meV) | $^{12}C^{18}O$ (meV) |
|----|---|---|---|
| FR[001] | 35.54 | 34.26 (-3.6%) | 34.96 (-1.6%) |
| FR[$\bar{1}10$] | ≈18 | ≈18 | ≈18 |
| FT[$\bar{1}10$] | 12.33 | 12.13 (-1.6%) | 11.68 (-5.3%) |
| FT[001] | ≈8 | ≈8 | ≈8 |

**Table S2. Theoretical investigation on the low-energy vibrational energy shifts for an isotope substituted CO molecule with the Cu$_{11}$ tip.** (**a**) The case of CO on a top site is shown, where the tip is located on the top site far from the surface. The numbers in the table represent the vibrational energies of CO (in parenthesis, the isotope shift relative to the normal molecule). (**b-c**) The same as (a), but for the case of the tip located very close to the CO molecule, where the CO molecule on the bridge is the most stable geometry.

**a**, CO on top with Cu$_{11}$ tip: $z_{cal}$=600 pm

|  | $^{12}C^{16}O$ (meV) | $^{13}C^{16}O$ (meV) | $^{12}C^{18}O$ (meV) |
|---|---|---|---|
| FR[$\bar{1}$10] | 31.982 | 30.964 (-3.2%) | 31.654 (-1.0%) |
| FR[001] | 29.519 | 28.568 (-3.2%) | 29.231 (-1.0%) |
| FT[$\bar{1}$10] | 4.491 | 4.457 (-0.8%) | 4.278 (-4.7%) |
| FT[001] | 4.047 | 4.018 (-0.7%) | 3.854 (-4.8%) |

**b**, CO on bridge with Cu$_{11}$ tip: $z_{cal}$=100 pm

|  | $^{12}C^{16}O$ (meV) | $^{13}C^{16}O$ (meV) | $^{12}C^{18}O$ (meV) |
|---|---|---|---|
| FR[001] | 33.016 | 31.962 (-3.2%) | 32.513 (-1.0%) |
| FR[$\bar{1}$10] | 24.416 | 23.574 (-3.4%) | 24.263 (-0.6%) |
| FT[$\bar{1}$10] | 16.669 | 16.602 (-0.4%) | 15.789 (-5.3%) |
| FT[001] | 8.294 | 8.232 (-0.7%) | 7.900 (-4.8%) |

**c**, CO on bridge with Cu$_{11}$ tip: $z_{cal}$=90 pm

|  | $^{12}C^{16}O$ (meV) | $^{13}C^{16}O$ (meV) | $^{12}C^{18}O$ (meV) |
|---|---|---|---|
| FR[001] | 32.846 | 31.799 (-3.2%) | 32.513 (-1.2%) |
| FR[$\bar{1}$10] | 25.450 | 24.541 (-3.6%) | 25.318 (-0.5%) |
| FT[$\bar{1}$10] | 17.036 | 17.000 (-0.2%) | 16.101 (-5.5%) |
| FT[001] | 8.517 | 8.453 (-0.8%) | 8.112 (-4.8%) |

**Table S3. Theoretical investigation on the low-energy vibrational energy shifts for an isotope substituted CO molecule with the Cu$_5$ tip.** (**a-c**) The same as Table S2 but for the Cu$_5$ tip.

**a**, CO on top with Cu$_5$ tip: $z_{cal}$=600 pm

|  | $^{12}C^{16}O$ (meV) | $^{13}C^{16}O$ (meV) | $^{12}C^{18}O$ (meV) |
|---|---|---|---|
| FR[$\bar{1}$10] | 31.967 | 30.949 (-3.2%) | 31.640 (-1.0%) |
| FR[001] | 29.517 | 28.556 (-3.3%) | 29.230 (-1.0%) |
| FT[$\bar{1}$10] | 4.519 | 4.485 (-0.8%) | 4.305 (-4.7%) |
| FT[001] | 4.020 | 3.991 (-0.7%) | 3.827 (-4.8%) |

**b**, CO on bridge with Cu$_5$ tip: $z_{cal}$=100 pm

|  | $^{12}C^{16}O$ (meV) | $^{13}C^{16}O$ (meV) | $^{12}C^{18}O$ (meV) |
|---|---|---|---|
| FR[001] | 33.235 | 32.212 (-3.1%) | 32.847 (-1.2%) |
| FR[$\bar{1}$10] | 22.170 | 21.570 (-2.7%) | 21.800 (-1.7%) |
| FT[$\bar{1}$10] | 15.173 | 15.030 (-0.9%) | 14.470 (-4.6%) |
| FT[001] | 10.592 | 10.499 (-0.9%) | 10.104 (-4.6%) |

**c**, CO on bridge with Cu$_5$ tip: $z_{cal}$=90 pm

|  | $^{12}C^{16}O$ (meV) | $^{13}C^{16}O$ (meV) | $^{12}C^{18}O$ (meV) |
|---|---|---|---|
| FR[001] | 32.994 | 31.980 (-3.1%) | 32.608 (-1.2%) |
| FR[$\bar{1}$10] | 23.818 | 23.019 (-3.4%) | 23.621 (-0.8%) |
| FT[$\bar{1}$10] | 16.282 | 16.224 (-0.4%) | 15.415 (-5.3%) |
| FT[001] | 10.601 | 10.508 (-0.9%) | 10.112 (-4.6%) |

**Table S4. Investigation of the appropriate exchange-correlation functional to describe the interaction between a CO molecule and a Cu(110) surface.** For each of the considered functionals the computed lattice constant ($a_0$) for the copper crystal is listed together with the corresponding adsorption energies for four adsorption sites of CO on Cu(110) (top, bridge, low-top and low-bridge sites). The most stable configuration is highlighted in boldface. Consistent with Ref. 51, only the vdw-DF and vdW-DF2 functionals correctly predict that CO adsorbs on a top site, where the adsorption energies are close to the experimental value of 0.63 eV [52]. On the other hand, $a_0$ for these functionals are a bit larger than the experimental value of 361 pm as discussed in Ref. 53.

| Functional | $a_0$ (pm) | Top (eV) | Bridge (eV) | low-top (eV) | low-bridge (eV) |
|---|---|---|---|---|---|
| PBE | 364 | -0.918 | **-0.987** | +0.053 | +0.010 |
| vdW-optB86b | 360 | -1.059 | **-1.142** | -0.109 | -0.128 |
| vdW-DF (revPBE-vdW) | 371 | **-0.681** | -0.608 | +0.045 | -0.054 |
| vdW-DF2 (rPW86-vdW2) | 375 | **-0.627** | -0.520 | +0.063 | -0.185 |